# Ultrafast Magnetic Switching of GdFeCo with Electronic Heat Currents

R. B. Wilson[1,*], Jon Gorchon[2,3,*], Yang Yang[4], Charles-Henri Lambert[2],

Sayeef Salahuddin[2,3], Jeffrey Bokor[2,3].

1) Department of Mechanical Engineering and Materials Science & Engineering Program, University of California, Riverside, CA 92521, USA

2) Department of Electrical Engineering and Computer Sciences, University of California, Berkeley, CA 94720, USA

3) Lawrence Berkeley National Laboratory, 1 Cyclotron Road, Berkeley, CA 94720, USA

4) Department of Materials Science and Engineering, University of California, Berkeley, CA 94720, USA

correspondence should be addressed to rwilson@engr.ucr.edu and jbokor@berkeley.edu.

*denotes equal contribution

## Abstract

We report the magnetic response of Pt/Au/GdFeCo trilayers to optical irradiation of the Pt surface. For bilayers with Au thickness greater than 50 nm, the great majority of energy is absorbed by the Pt layer, creating an initial temperature differential of thousands of Kelvin between the Pt/Au layers and the GdFeCo layer. The resulting electronic heat current across the metal multilayer lasts for several picoseconds with energy flux in excess of 2 TW $m^{-2}$ and provides sufficient heating to the GdFeCo electrons to induce deterministic reversal of the magnetic moment.

# I. Introduction

Ultrafast reversal of the magnetic moment can be optically induced in metals that possess two antiferromagnetically coupled sublattices, e.g. Gd and FeCo [1]. All optical switching (AOS), was first observed by Stanciu *et al.* in 2007 [2]. In AOS experiments, GdFeCo conduction electrons are excited with an ultrafast laser pulse to eV energies above the Fermi level [3]. Subsequently, the FeCo sublattice demagnetizes within a few hundred femtoseconds [1]. The Gd sublattice also loses magnetic order, but at a slower rate [1]. The differing rates of demagnetization, together with the transfer of angular momentum from the Gd to FeCo sublattice, enables reversal of the magnetic moment on ps time-scales [3-5]. While initial studies credited the ultrafast reversal of the magnetization to a helicity-dependent light-matter interaction [2], subsequent investigations with linearly polarized light demonstrate the reversal is driven solely by energy absorption [4,6].

Here, we demonstrate direct laser irradiation of GdFeCo is not necessary for deterministic reversal of the magnetization. Purely electronic heat currents are also effective at switching. Our work, which focuses on how indirect excitation of a ferrimagnetic metal impacts ultrafast switching, builds on several recent studies of how indirect excitation of ferromagnetic metals impacts magnetization dynamics [7-10]. Our work also builds on recent experimental investigations into the role of temperature on all optical switching phenomena [4,11,12]. We report the magnetic response of 5 nm Pt/ *h* nm Au/ 10 nm GdFeCo trilayers to optical irradiation at the Pt surface (Fig. 1). By varying the Au thickness *h* from 0 to 200 nm, we control the ratio of laser energy directly absorbed by the GdFeCo vs. the Pt and Au layers. The total fluence that needs to be absorbed by the trilayer to cause the GdFeCo magnetization to reverse increases by only a factor of eight when the Au thickness is increased from 0 to 200 nm, despite a negligible amount of energy being directly absorbed by the GdFeCo when the Au film is thick (Fig 2). Our results demonstrate electronic heat currents can reverse the magnetization as efficiently as direct optical irradiation.

## II. Experimental Methods

We focus our study on six Pt/Au/GdFeCo trilayer samples prepared via magnetron sputter deposition on sapphire substrates. The Au film thicknesses for the six samples are 0, 10, 30, 72, 113, and 190 nm. The GdFeCo film thickness is ~10 nm in all six samples. The Pt film thickness is ~6 nm in all samples. Layer thicknesses are based on a combination of X-ray reflectivity measurements of the multilayers with a total thickness less than 100 nm, and calibrated sputter deposition rates for thicker samples. The GdFeCo films were prepared via co-sputtering of a Gd and $Fe_{90}Co_{10}$ target. Based on calibrated sputter deposition rates of the Gd and $Fe_{90}Co_{10}$ targets, we estimate the $Gd_x(Fe_{90}Co_{10})_{1-x}$ composition to be $x = 0.34$. The compensation temperature of the GdFeCo films is below room temperature.

We use an amplified Ti:sapphire laser with 810 nm center wavelength in our experiments (Coherent RegA 9050). The laser pulse duration full-width at half maximum (FWHM) is 55 fs. We run the laser amplifier at a repetition rate of 250 kHz for time-resolved magneto-optic Kerr effect (MOKE) measurements, or instead eject single laser pulses for MOKE micrograph imaging of single-shot switching.

We use a MOKE microscope for monitoring the GdFeCo magnetization after laser irradiation with single laser pulses (Fig. 1b). The MOKE microscope focuses on the GdFeCo film through the sapphire substrate. In these experiments, an external magnetic field H ≈ ±100 Oe saturates the magnetization of the sample out-of-plane. Following removal of the external field, a single linearly polarized laser pulse irradiates the Pt surface. As shown in Fig. 1b for the Pt/Au 113 nm/ GdFeCo sample, if a laser pulse of sufficient energy irradiates the Au film surface, the magnetization of the irradiated region reliably toggles between white (up) and black (down).

We also performed time-resolved pump-probe MOKE measurements on the samples to investigate the ultrafast magnetization dynamics of the FeCo sublattice following laser irradation, see Fig. 1c. In these experiments, the pump laser is incident on the Pt surface of the trilayer, while the probe laser is focused on the surface of the GdFeCo film, through the sapphire substrate. The pump beam $e^{-2}$ radius is 65 µm. The probe beam $e^{-2}$ radius is ~10 µm. During time-resolved magneto optic Kerr effect measurements, a constant perpendicular field of ~50 Oe is applied to reset the magnetization between pump pulses.

## III. **Results and Analysis**

In all samples, regardless of Au film thickness, irradiation of the Pt surface with sufficient fluence causes an observable reversal of the GdFeCo magnetization. In Fig. 2, we report the total fluence, $F_T$, the sample must absorb to induce magnetization reversal of the GdFeCo. We calculate the total absorption, and relative absorption in each layer using a multilayer optical calculation [11]. For samples with Au films thicker than 30 nm, a negligible amount of optical energy is directly absorbed by the GdFeCo film. Therefore, we conclude that for the samples with Au layers thicker than 30 nm, electronic heat currents flowing from the adjacent Au layer are responsible for the deterministic reversal of the GdFeCo magnetic moment.

To confirm that electronic heat-currents are responsible for switching, we performed a control experiment on a Pt (5 nm)/Au(75 nm)/MgO(3 nm)/Au(5 nm)/GdFeCo(10 nm) sample. For this sample, the insulating MgO layer prevents electronic heat currents into the GdFeCo. No magnetization reversal is observed in this sample at any fluence.

To interpret our experimental data, we use a thermal model to predict the temperature responses of the electrons and phonons in the trilayers, see Fig. 3. Our thermal model is a multilayer variation of the well-known “two-temperature” model and consists of two coupled heat diffusion equations for the electrons and phonons in each metal layer [13,14]. In the GdFeCo layer, we also add a third heat-equation to account for the ability of the spins in the GdFeCo layer to act as a thermal reservoir [11,15]. The electron heat-diffusion equation includes a heat generation term to account for the optical energy deposited through the depths of the multilayer. The depth dependence of the absorption is calculated using a multilayer optical calculation. We solve the coupled heat-diffusion equations numerically via a Crank-Nicolson finite difference method. In prior work, we have used a variation of the two-temperature thermal model we use here to quantitatively describe heat transfer across Pt/Au bilayers [13], ultrafast demagnetization of FePt:Cu thin films [16], and investigate the role of electron and phonon temperatures in all-optical switching of GdFeCo [11]. Further details of the thermal model are contained in in Refs. [11,13,14,16,17]. We emphasize that all the thermal properties of the system are fixed based on the results of prior publications on transport in Pt/Au bilayers [13], and a prior study of GdFeCo thin films [11]. No model parameters are adjusted to improve agreement between the model predictions and experimental results.

Following optical heating of the Pt layer, our thermal model predicts the average temperature of the Au electrons exceeds 1000 K (Fig 3a). The high diffusivity of the Au electrons allows rapid heat diffusion [13], resulting in TW $m^{-2}$ picosecond heat currents into the GdFeCo (Fig. 3b) in samples with thick Au films despite negligible direct optical absorption. Electronic charge currents play no role in our experiments because the dielectric relaxation time in metals is on the order of $10^{-18}$ s [18].

The Pt/Au/GdFeCo trilayers with Au layers require more energy to be absorbed by the Pt layer (Fig. 1) because only a fraction of the energy absorbed by the Pt and Au layers diffusing across the Au layer into the GdFeCo electrons. In parallel to energy transfer from the hot Au electrons to the GdFeCo electrons, significant energy is transferred to the Au phonons via electron-phonon scattering [13]. The characteristic length-scale at room temperature over which the electronic heat can diffuse before the hot Au electrons transfer most of their energy to the phonons is $d_{ep} \approx \sqrt{\Lambda_{e,Au} / g_{ep,Au}} \approx 100\ \mathrm{nm}$, where $\Lambda_{e,Au} \approx 250\ \mathrm{W\,m^{-1}\,K^{-1}}$ is the thermal conductivity of the Au electrons at room temperature and $g_{ep,Au} \approx 2.2\times10^{16}\ \mathrm{W\,m^{-3}\,K^{-1}}$ is the electron-phonon coupling constant of Au [13]. The electronic thermal conductivity is proportional to the electronic heat-capacity, and total scattering rate from defects and phonons, $\Lambda_e = C_e v_f^2 \tau / 3$. At high electron temperatures, the thermal conductivity increases due to an increase in electronic heat capacity. The scattering rate from defects remains unchanged at high temperatures, while the change in scattering rate from phonons is small because the change in phonon temperature is small for the first few picoseconds of the experiment. In our experiments, we expect that $\Lambda_{e,Au}$ will exceed $10^3\ \mathrm{W\,m^{-1}\,K^{-1}}$ on picosecond time-scales due to the high electron temperatures, corresponding to a $d_{ep}$ of more than 200 nm for ~ 2 picoseconds following laser irradiation.

In samples where the Au layer is greater than 30 nm, the GdFeCo electrons are only heated indirectly through electronic heat-currents from the adjacent Au layer. Integrating both the optical and electronic heat-currents over the time interval of the experiment yields the total fluence absorbed by the GdFeCo electrons, $F_{GFC}$, as a function of Au thickness (Fig. 4). We observe in all samples that a total fluence between 5 and 6 J $m^{-2}$ must be absorbed by the GdFeCo from electronic and/or optical heat currents for magnetization reversal to occur.

In our thermal analysis, above, we assume the electrons transport heat to the GdFeCo layer diffusively. However, the laser initially excites a nonthermal distribution of electrons. Therefore, ballistic or superdiffusive transport is also theoretically possible on sub picosecond time-scales. The conditions necessary for ballistic vs. diffusive transport in nanoscale metal multilayers is an active area of research, and no consensus currently exists. For example, Choi *et al*. report pump/probe thermoreflectance measurements of 80 nm thick Pt/Au bilayers that are consistent with diffusive transport, regardless of whether the Pt or Au layer is irradiated [13]. Alternatively, several pump/probe measurements have examined the time-scale for energy to diffuse across Cu or Au films that are hundreds of nanometers thick and concluded transport is ballistic on sub-picosecond time-scales in these materials [7,19].

To examine whether transport is predominantly ballistic or diffusive in the present experiments, we consider the time-scale for energy to traverse across samples of different thickness. The time-scale for energy to ballistically traverse 75, 120, and 200 nm thick Au layers is given by $\tau \approx h_{Au}/v_F \approx$ 50, 85, and 140 fs, respectively. Here, $v_F \approx 1.4\times10^6$ m/s is the Fermi velocity of Au. Alternatively, the time-scales for diffusive transport (Fig. 3) are 0.4, 0.8, and 1.4 ps for the samples with 75 120, and 200 nm thick Au layers. We note that the time-scale for diffusive energy transport in a metal following laser irradiation is not related to the speed of sound of the metal [7,19], but is instead determined by the thermal diffusivity of the hot electrons [13,20].

To experimentally examine the time-scale for energy to traverse across Au layers of different thickness, we performed time-resolved MOKE measurements on the samples with Au film thickness of 75, 120, and 200 nm with a fixed incident fluence, see Fig. 5. The laser fluence incident on the Pt layer in these measurements is ~100 J m$^{-2}$, corresponding to an absorbed fluence of ~18 J m$^{-2}$. Included in Fig. 5 for comparison is time-resolved MOKE data or the Pt/GdFeCo sample with an incident fluence of 14 J m$^{-2}$. An incident fluence of 14 J m$^{-2}$ on the Pt/GdFeCo sample corresponds to ~2.2 and 2.8 J m$^{-2}$ of fluence absorbed by the Pt and GdFeCo layers, respectively.

Prior studies indicate that the magnetization of GdFeCo responds to heating of the electrons on time-scales of less than 0.2 ps, comparable to the time-scale for ballistic transport across a few hundred nm of Au. Therefore, if transport across the Au layer were ballistic, demagnetization of all samples should occur within the first few hundred femtoseconds of laser irradiation. Instead, we observe a substantial delay in demagnetization in the samples with thick Au layers relative to the sample with no Au layer, see Fig. 5. In the samples with 72, 113, and 200 nm thick Au layers, the lag in demagnetization in comparison to the sample with no Au film is 0.65, 0.65, and 0.95 ps. Here, we define the demagnetization time-scale as the delay time where demagnetization is 50% of its peak value, i.e. the delay time where $\Delta M / M_s$ reaches 0.07, 0.11, and 0.21 for the samples with 72, 113, and 200 nm thick Au films.

Another test for whether energy transport is ballistic or diffusive is the quantity of energy that reaches the GdFeCo layer in the first few picoseconds. The mean-square-displacement of energy increases quadratically with time for ballistic transport, but only linearly with time for diffusive transport. Mean-square displacement is a measure of how energy is spatially dispersed. The data in Fig. 5 provides an estimate of the amount of the energy that reaches the GdFeCo. Ten picoseconds following absorption of 18 J m$^{-2}$ in the Pt and Au layers, $\Delta M / M_s$ = 0.3, 0.18, and 0.09 for the samples with 75, 120, and 200 nm thick Au films, see Fig. 5. Pump-probe measurements of the Pt/GdFeCo sample with no Au film as a function of fluence indicate that to induce $\Delta M / M_s$ = 0.3, 0.18, and 0.09 requires the GdFeCo layer absorb fluences of 3.5, 2.6, and 1.6 J m$^{-2}$. Therefore, by dividing these values by the 18 J m$^{-2}$ optically absorbed by the Pt/Au layers, we conclude that the energy transmission across the 75, 120, and 200 nm thick Au films is 19, 14, and 9%. These values are consistent with our thermal model. Our thermal predicts that 22, 15, and 7% of the fluence initially absorbed by the Pt and Au electrons will traverse Au film thicknesses of 72, 113, and 200 nm and reach the GdFeCo layer within the first 5 ps.

Our demonstration that electronic heat-currents can induce magnetization reversal provides an important experimental test of the role of thermal vs. nonthermal electrons in ultrafast magnetic switching [21-28]. To date, there has been a mismatch between experimental and theoretical studies. Prior theoretical studies of the switching phenomena assume the initial distribution of excited electrons is thermal. In contrast, prior experimental studies of all optical switching have used optical irradiation to excite nonthermal distributions of electrons that are not well described by Fermi-Dirac statistics [29]. Several reasons exist to believe an initially non-thermal distribution of electrons can impact the magnetization dynamics. Highly excited nonthermal electrons could allow for magnetization quenching via the generation of Stoner excitations [27,29]. Nonthermal distributions enable nonlocal superdiffusive transport of energy and angular momentum [22,23,25]. Finally, the magnitude and duration of energy transfer between electrons and phonons in a metal depends strongly on whether the initial distribution of electrons is thermal or nonthermal [20]. Experimental [16,30] and theoretical studies [31] demonstrate that the rate of energy exchange between electrons and phonons can dramatically impact the dynamics of either ferromagnetic or ferrimagnetic metals. Our results demonstrate that while the types of nonthermal phenomena described above may play a secondary role in all optical switching, they are not required for switching to occur, which is consistent with theoretical modelling of the phenomena [4,31].

In conclusion, by adding a Pt/Au bilayer adjacent to GdFeCo to serve as an optical absorber, we examine how exciting GdFeCo with electronic thermal currents differs from direct optical excitation. We observe that excitation of GdFeCo with picosecond electronic heat-currents also induces a reversal of the magnetization of GdFeCo magnetic layers. The discovery that electronic heat currents are effective in magnetization reversal of GdFeCo signals new opportunities for potential device applications of ultrafast magnetization switching.

**Acknowledgements**

This work was primarily supported by the Director, Office of Science, Office of Basic Energy Sciences, Materials Sciences and Engineering Division, of the U.S. Department of Energy under Contract No. DE-AC02-05-CH11231 within the Nonequilibrium Magnetic Materials Program (MSMAG). We also acknowledge the National Science Foundation Center for Energy Efficient Electronics Science for providing most of the experimental equipment and financially supporting Y.Y. during his work on this project.

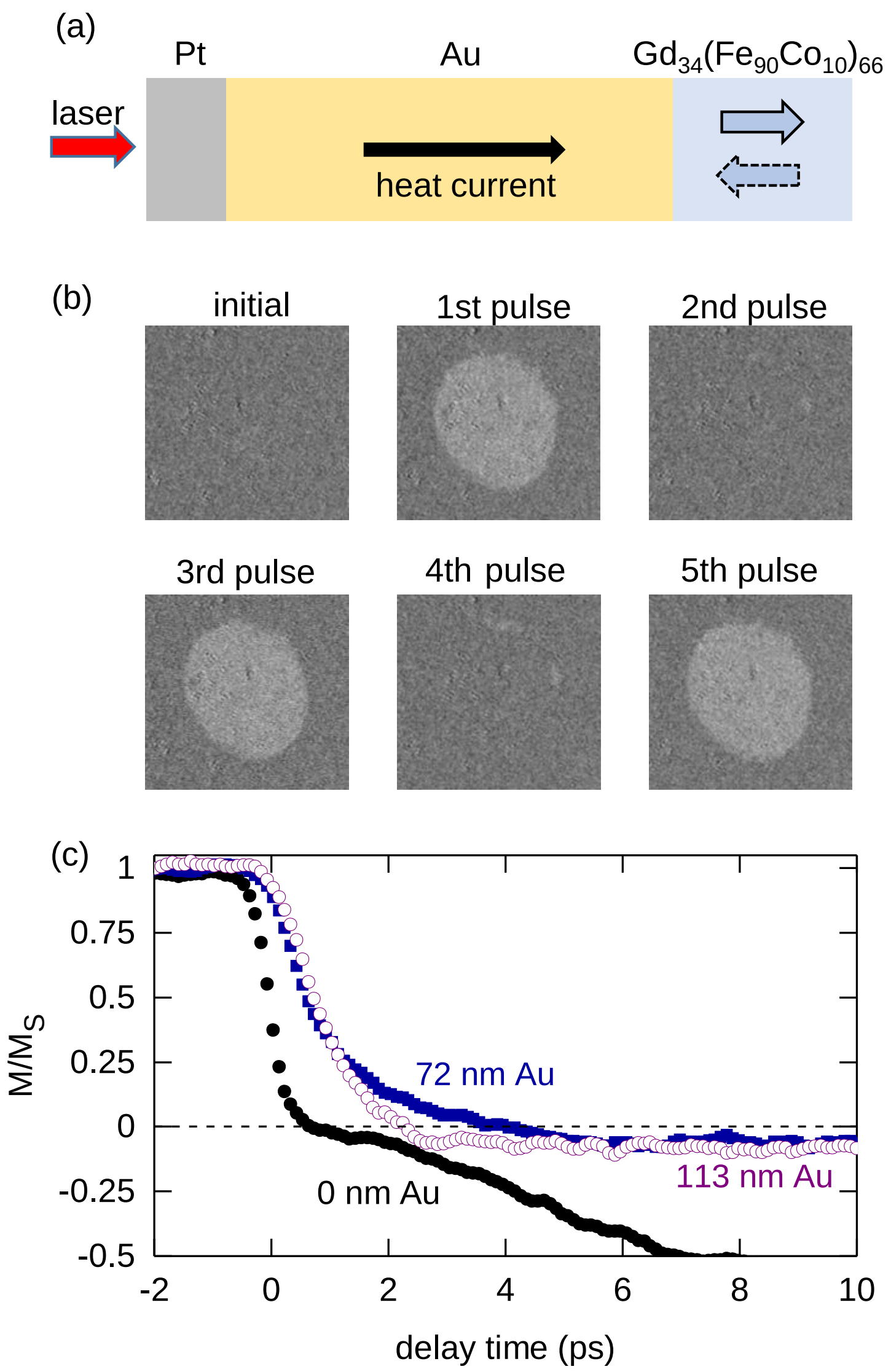

Fig. 1. (a) Schematic of experiment. A laser pulse irradiates the Pt surface and deposits energy in the Pt and Au electrons. Hot electrons diffuse across the Au layer and heat the GdFeCo. (b) MOKE micrographs of the GdFeCo magnetization in a 6 nm Pt/ 113 nm Au / 10 nm GdFeCo trilayer after the Au surface is successively irradiated with linearly polarized laser pulses. The sample's initial magnetization is down (M−). (c) Time-resolved MOKE data of magnetization switching following incident irradiation of the samples with 0, 72, and 113 nm thick Au films with 25, 150 and 240 J m$^{-2}$.

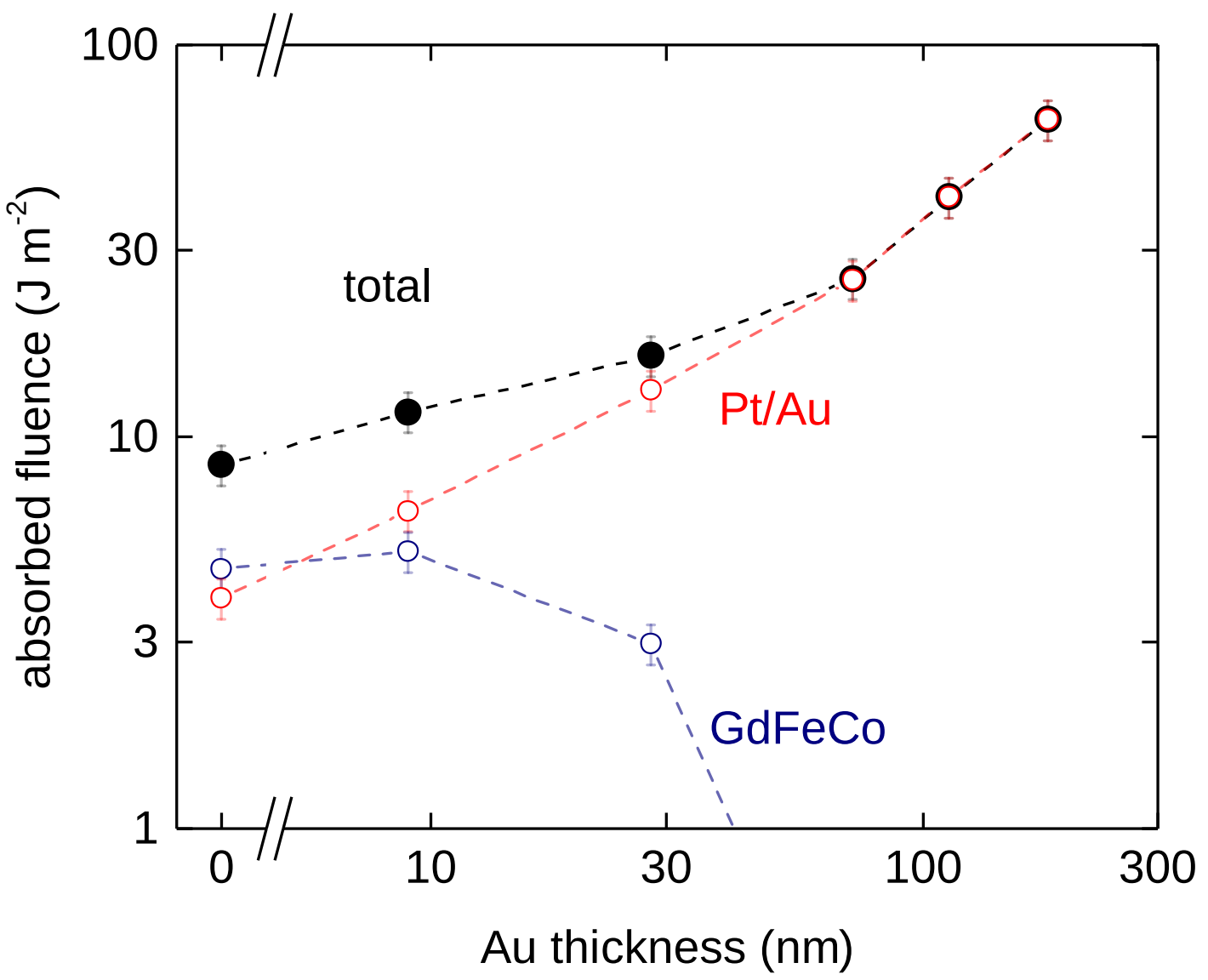

Fig. 2. Dependence of the absorbed laser fluence required for reversing the magnetization of the GdFeCo as function of the thickness of the Au layer. Lines are to guide the eye. A multilayer optical calculation with $n = 2.85 + 5i$ for Pt [13,32], $n = 0.2 + 4.9i$ for Au [13,32], and $n = 3.2 + 3.5i$ for GdFeCo [11] determines the amount of fluence absorbed in the GdFeCo layer vs. the Pt and Au layers for each sample.

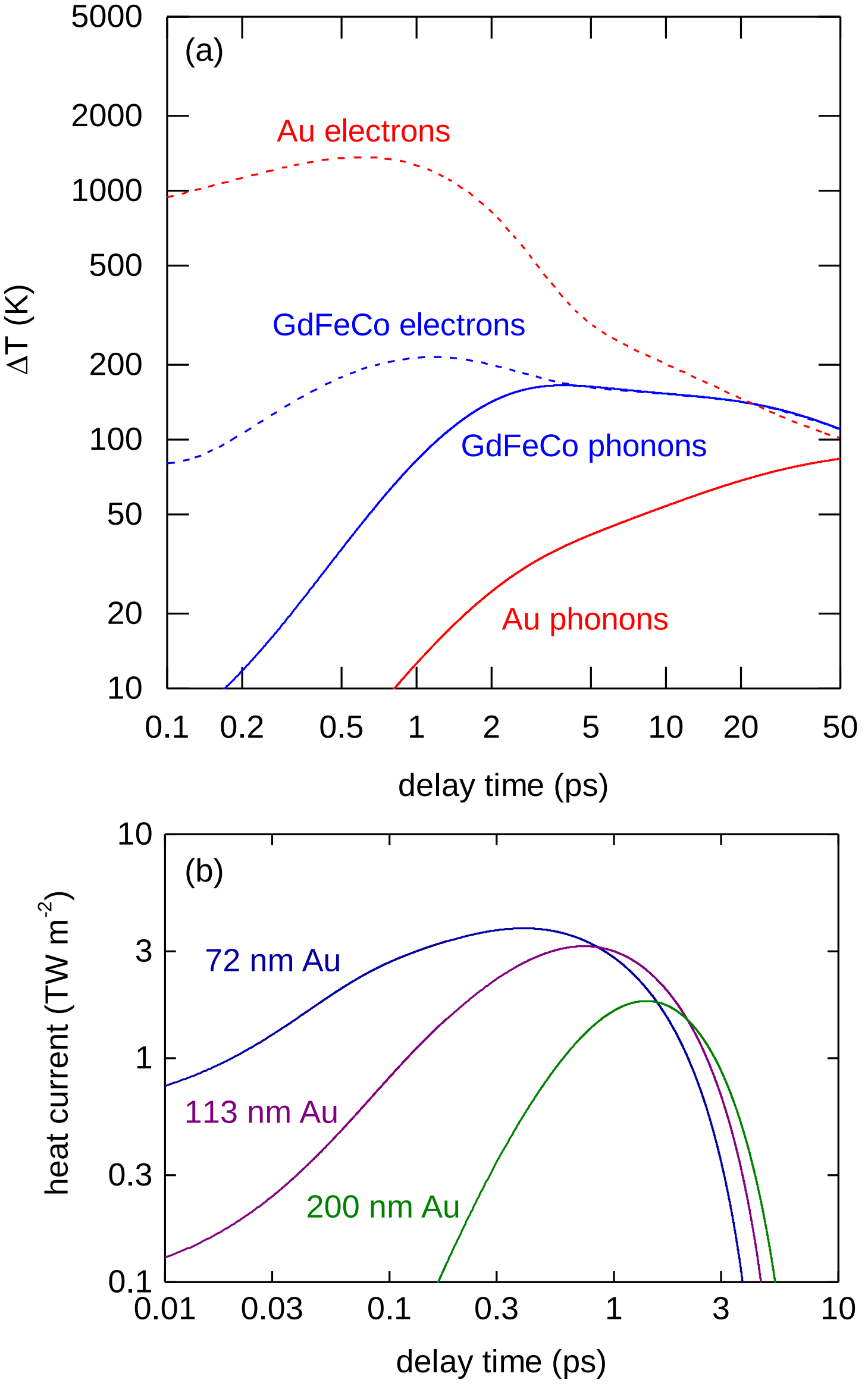

Figure 3. (a) Temperature response of the Pt (6 nm) / Au (72 nm) / GdFeCo (10 nm) trilayer after the Pt and Au electrons absorb 27 J $m^{-2}$ from a 55 fs laser pulse. Each curve represents an average temperature across the layer. The large temperature difference between the Au electrons and GdFeCo electrons for the first few picoseconds following irradiation generates large electronic heat currents. (b) Heat currents into the GdFeCo electrons via hot electrons from the adjacent Au film. The 4heat-currents shown for the samples with 72, 113, and 200 nm thick Au layers correspond to total absorbed fluences in the Pt and Au layers of 27, 44, and 70 J $m^{-2}$.

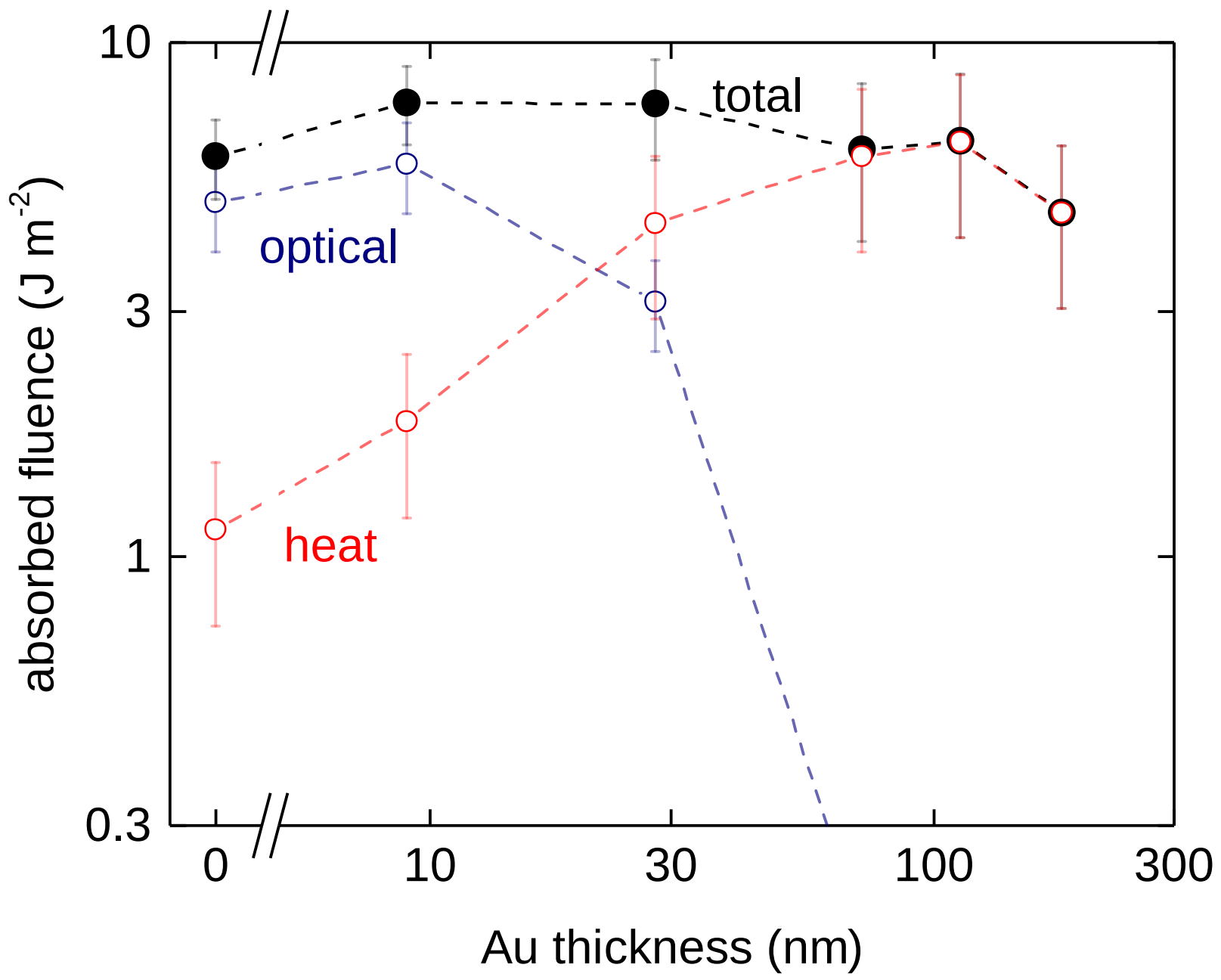

Figure 4. Fluence absorbed by GdFeCo electrons vs. Au film thickness. The red open circles demark the fluence from heat currents via the Au electrons, the blue open circles represent the fluence from direct optical absorption, and the filled black circles represent the total fluence absorbed by the GdFeCo electrons from all sources. Lines are to guide the eye.

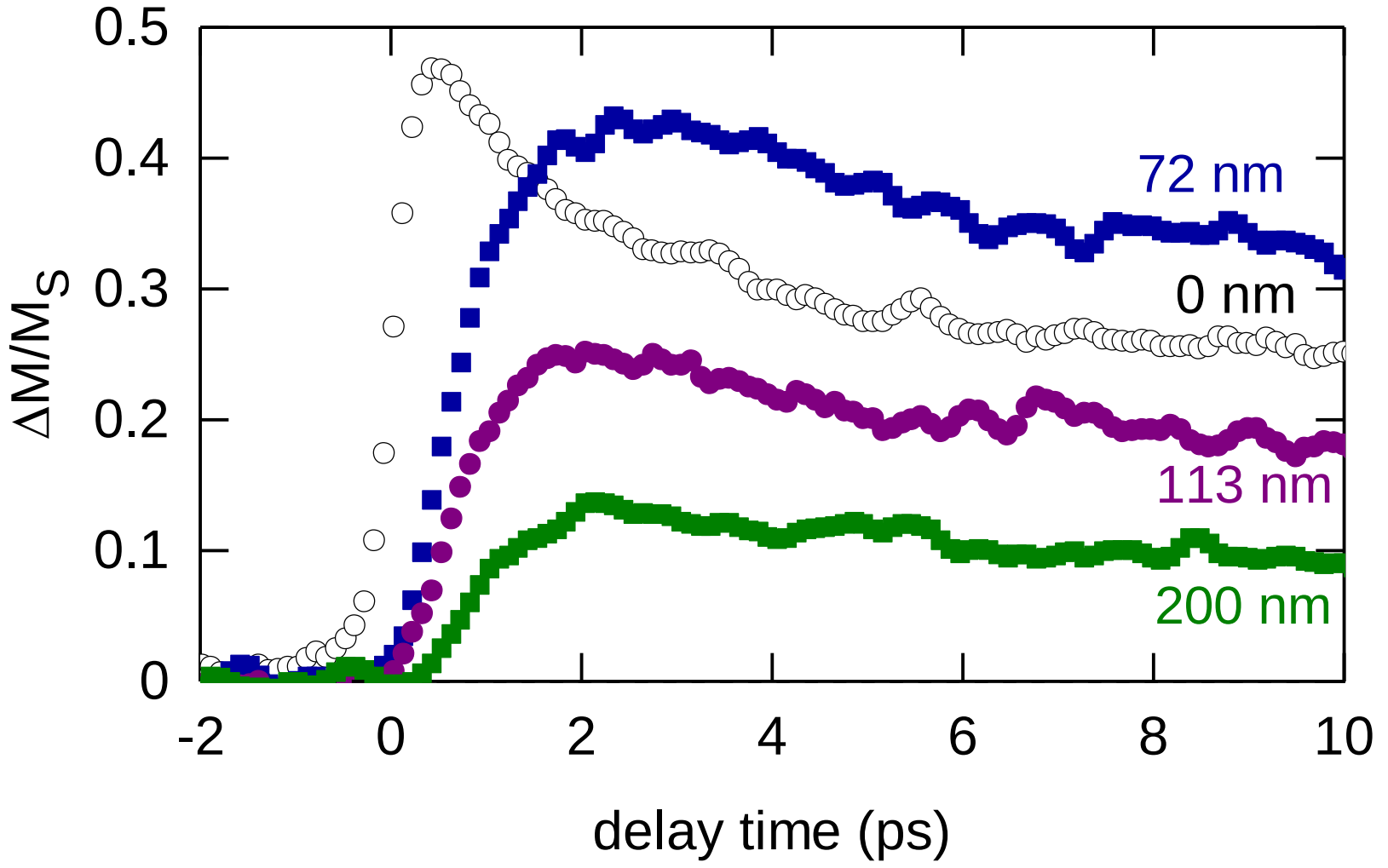

Fig. 5. Time-resolved MOKE measurements of Pt/Au/GdFeCo samples. The pump laser is incident on the Pt. The incident fluence on the sample with no Au layer is 14 J m$^{-2}$, while the incident fluence on the other three samples is ~100 J m$^{-2}$. The 0.65, 0.65 and 0.95 ps delay in demagnetization of the samples with 72, 113, and 200 nm thick Au layer between the Pt absorber and GdFeCo layer is consistent with diffusive heat transfer by hot Au electrons.